\documentclass[lettersize,journal]{IEEEtran}
\usepackage{amsmath,amsfonts}
\usepackage{algorithm}
\usepackage{array}
\usepackage[caption=false,font=normalsize,labelfont=sf,textfont=sf]{subfig}
\usepackage{textcomp}
\usepackage{stfloats}
\usepackage{url}
\usepackage{verbatim}
\usepackage{graphicx}
\usepackage{cite}
\usepackage{amsmath,amssymb,amsfonts}
\usepackage{graphicx}
\usepackage{xcolor}
\usepackage{physics}
\usepackage{gensymb}
\usepackage{longtable}
\usepackage{hyperref}
\usepackage{expl3}
\usepackage{algpseudocode}

\begin{document}
\title{A DSP-Free Carrier Phase Recovery System using 16-Offset-QAM Laser Forwarded Links for 400Gb/s and Beyond}

\author{Marziyeh~Rezaei,~\IEEEmembership{Student,~IEEE},
        Dan~Sturm,~\IEEEmembership{Student,~IEEE},
        Pengyu~Zeng,~\IEEEmembership{Student,~IEEE},
        and~Sajjad~Moazeni,~\IEEEmembership{Member,~IEEE}}

\markboth{Journal of \LaTeX\ Class Files,~Vol.~14, No.~8, August~2021}%
{Shell \MakeLowercase{\textit{et al.}}: A Sample Article Using IEEEtran.cls for IEEE Journals}

\maketitle
\begin{abstract}
Optical interconnects are becoming a major bottleneck in scaling up future GPU racks and network switches within data centers. Although 200\,Gb/s optical transceivers using PAM-4 modulation have been demonstrated, achieving higher data rates and energy efficiencies requires high-order coherent modulations like 16-QAM. Current coherent links rely on energy-intensive digital signal processing (DSP) for channel impairment compensation and carrier phase recovery (CPR), which consumes approximately 50\,pJ / b - 10$\times$ higher than future intra-data center requirements. For shorter links, simpler or DSP-free CPR methods can significantly reduce power and complexity. While Costas loops enable CPR for QPSK, they face challenges in scaling to higher-order modulations (e.g., 16/64-QAM) due to varying symbol amplitudes.

In this work, we propose an optical coherent link architecture using laser forwarding and a novel DSP-free CPR system using offset-QAM modulation. 
The proposed analog CPR feedback loop is highly scalable, capable of supporting arbitrary offset-QAM modulations (e.g., 4, 16, 64) without requiring architectural modifications. This scalability is achieved through its phase error detection mechanism, which operates independently of the data rate and modulation type. We validated this method using GlobalFoundry’s monolithic 45nm silicon photonics PDK models, with circuit- and system-level implementation at 100GBaud in the O-band. We will investigate the feedback loop dynamics, circuit-level implementations, and phase-noise performance of the proposed CPR loop. Our method can be adopted to realize low-power QAM optical interconnects for future coherent-lite pluggable transceivers as well as co-packaged optics (CPO) applications.
\end{abstract}


\section{Introduction}
\label{Intro}
The rise of AI computing, driven by tens of thousands of GPUs and AI accelerators in data centers, poses significant challenges for optical interconnects and networking technologies~\cite{GenAIOptics-IEDM2023,9793366}. Current state-of-the-art optical interconnects utilize pluggable transceivers operating at 112\,Gb/s and 224\,Gb/s PAM-4 data rates with energy efficiencies of $\sim$25\,pJ/b~\cite{DSP-JLT2024}. Next-generation transceivers aim to achieve data rates exceeding 400\,Gb/s with improved energy efficiency~\cite{Radha-JLT2021}. However, with limited improvements in the speed and energy efficiency of mixed-signal circuitry in advanced CMOS technologies, and fiber dispersion constraints, further increases in data rates necessitate advanced modulation schemes. While PAM-6 and PAM-8 are under investigation, their significant SNR penalties~\cite{Li-OFC2019} make them less practical. Consequently, adopting advanced coherent modulation (e.g., 16-QAM) within data centers is inevitable for achieving energy-efficient links. Although wavelength and polarization division multiplexing (WDM and PDM) can increase aggregate bandwidth per fiber, boosting data rates per wavelength/polarization remains essential to reduce the number of laser lines and improve energy efficiency.

Coherent optical communication requires precise phase and frequency alignment between the received (Rx) signal and the local oscillator (LO). Long-haul coherent transceivers, such as the 800G-ZR/ZR+ standards~\cite{800GZR-JLT2023}, employ power-intensive digital signal processing (DSP) to handle channel impairments such as chromatic dispersion (CD), differential group delay (DGD), and carrier phase recovery (CPR)~\cite{DSP-JLT2024,CohDet-OptExp2008}. These DSP methods require oversampling ADCs, resulting in energy consumption around 50\,pJ/b and increased system complexity. In short-reach interconnects (e.g., inter-rack links $<$1\,km), many DSP functionalities—such as dispersion compensation—are unnecessary and can be eliminated~\cite{Saleh:21}. However, frequency and phase recovery remain essential. Therefore, performing CPR in analog front-end can eliminate or significantly simplify DSP, realizing the power efficiency requirements of data centers. 

Previous studies have demonstrated DSP-free CPR systems for QPSK/4-QAM modulations~\cite{ASP_USB_OPLL, ASP_USB_ODLL, ASP_IIT_PLL, ashok2021analog} using Costas loops~\cite{UCSB-Costas-OptExp2012, Wireless-Costas-2000}. These systems utilize optical phase-locked loops (OPLLs)~\cite{ASP_USB_OPLL}, optical delay-locked loops (ODLLs)~\cite{ASP_USB_ODLL}, and traditional PLLs~\cite{ASP_IIT_PLL} for phase and frequency mismatches in laser-forwarded links. However, as modulation schemes scale to higher orders (e.g., 16/64/256-QAM), Costas loops fail to correct phase errors due to varying symbol amplitudes~\cite{marziyeh_mwscas}. 

For instance, using unmodified Costas loops for 16-QAM introduces symbol-dependent phase error detection, creating false I/Q locking points that deviate from $\frac{n\pi}{2}$~\cite{QAM_16_CRP_wo_symbol_detection}. A proposed solution involves sub-sampling the 16-QAM signal to form a 4-QAM constellation~\cite{QAM_16_CRP_with_symbol_detection}, but this relies on accurate symbol detection during CPR, which is impractical. Furthermore, sub-sampling reduces CPR loop bandwidth. 

Another work~\cite{Pilot_tone_JLT} employed the pilot tone concept by adding a low-frequency test tone to the transmitted optical signal. At the receiver, phase error is detected and compensated based on the power of this tone. However, even after compensation, this low-power tone remains in the signal, interfering with the data and causing periodic jitter, ultimately impacting BER.

In this work, we propose a DSP-free carrier phase recovery system optimized for offset-QAM modulation. Offset-QAM is generated by leaking a portion of the LO signal into the modulated QAM signal at the transmitter, as shown in Fig.~\ref{Constellation}. This can be achieved using Mach-Zehnder Modulators (MZMs) operating with an offset from the transmission null~\cite{Lu:10}, or, as recently demonstrated, by employing micro-ring modulators (MRMs)~\cite{Dan_SUM}. In offset-QAM, the phase offset between the Rx and LO paths is detected based on the average power/voltage difference between the in-phase (I) and quadrature (Q) signals. This difference serves as an error signal to correct the phase error using a phase shifter (PS) in the LO path for laser-forwarded links. The laser-forwarding technique~\cite{JSSC2020-LaserFwrd} forwards a portion of the transmitter (Tx) laser to the receiver to serve as the LO, maintaining zero frequency mismatch, eliminating the need for an additional LO laser, and mitigating laser phase noise (PN). This approach reduces cost and enhances reliability. The proposed CPR system supports any offset-QAM modulation level (e.g., 4, 16, 256) without architectural modifications, making it highly scalable for future high-order modulations. Unlike Costas loops, our method avoids the need for high-speed analog devices (e.g., mixers, XORs) and does not suffer from $\frac{\pi}{2}$ phase ambiguity~\cite{Costas_loop_phase_ambiguity, Costas_loop_phase_ambiguity2}.

The proposed receiver is modeled and simulated using Global Foundry (GF) 45\,nm monolithic silicon photonics (GF45SPCLO)~\cite{GF45SPCLO-OFC2020} PDK models, with circuit/system-level implementation at 100\,GBaud. This approach paves the way for low-power QAM interconnects in future co-packaged optics (CPO) and coherent-lite~\cite{coh-lite-ieee2021} applications, enabling energy-efficient data rates exceeding 400\,Gb/s per channel.

The paper is organized as follows: Section~\ref{DSP-Free Techniques} introduces our DSP-free CPR techniques for offset-QAM modulation. Section~\ref{Loop Dynamics} analyzes loop dynamics, including stability and laser phase noise. Section~\ref{Circuit} discusses the circuit implementation. Section~\ref{sim results} presents simulation results and trade-offs. Finally, Section~\ref{conclusion} concludes the paper.

\section{Proposed DSP-Free Carrier Phase Recovery}
\label{DSP-Free Techniques}
\begin{figure}[tbp]
\noindent
  \centering{\includegraphics[width=\linewidth]{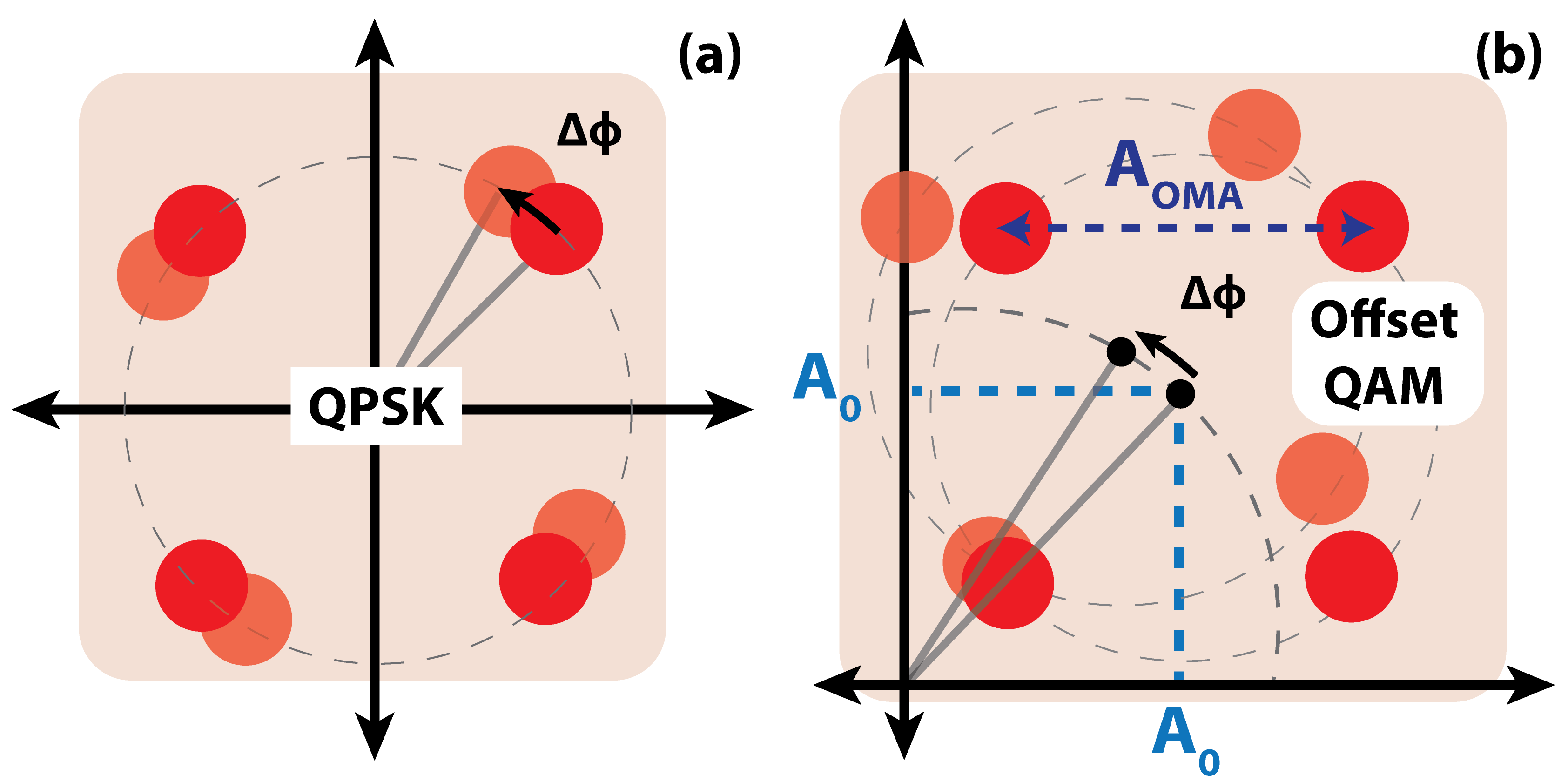}}
\caption{The constellation of (a) 4-QAM (QPSK) (b) 4-offset-QAM modulation with and without the phase error between the Rx and LO paths.}
\label{Constellation}
\end{figure}

\begin{figure}[tbp]
\noindent
  \centering
  {\includegraphics[width=80mm]{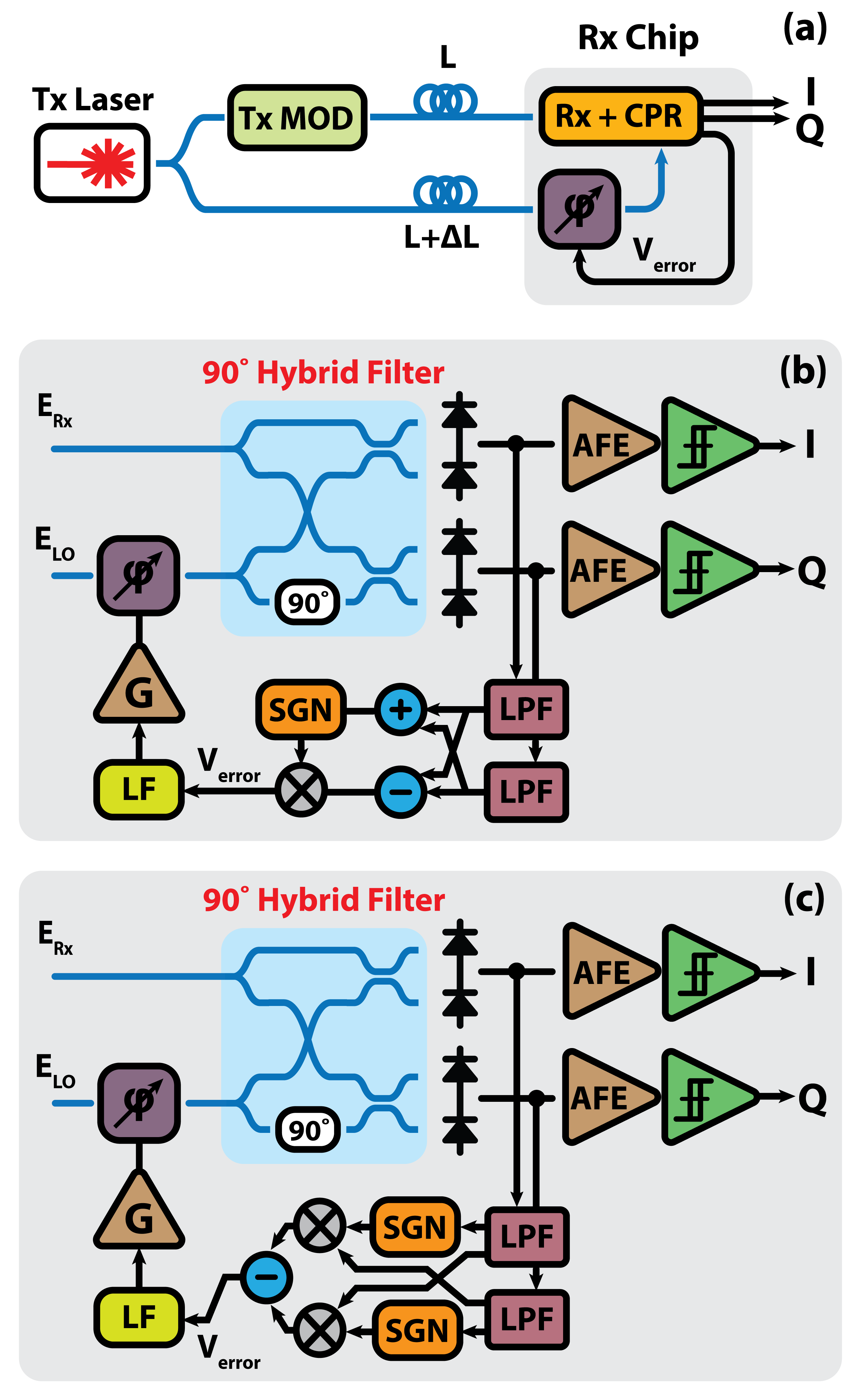}}
\caption{(a) Block diagrams of the proposed offset-QAM coherent receiver using Method 1 (b) and Method 2 (c) for LO phase recovery.}
\label{Methods}
\end{figure}

\begin{figure}[htbp]
\centerline{\includegraphics[width=\linewidth]{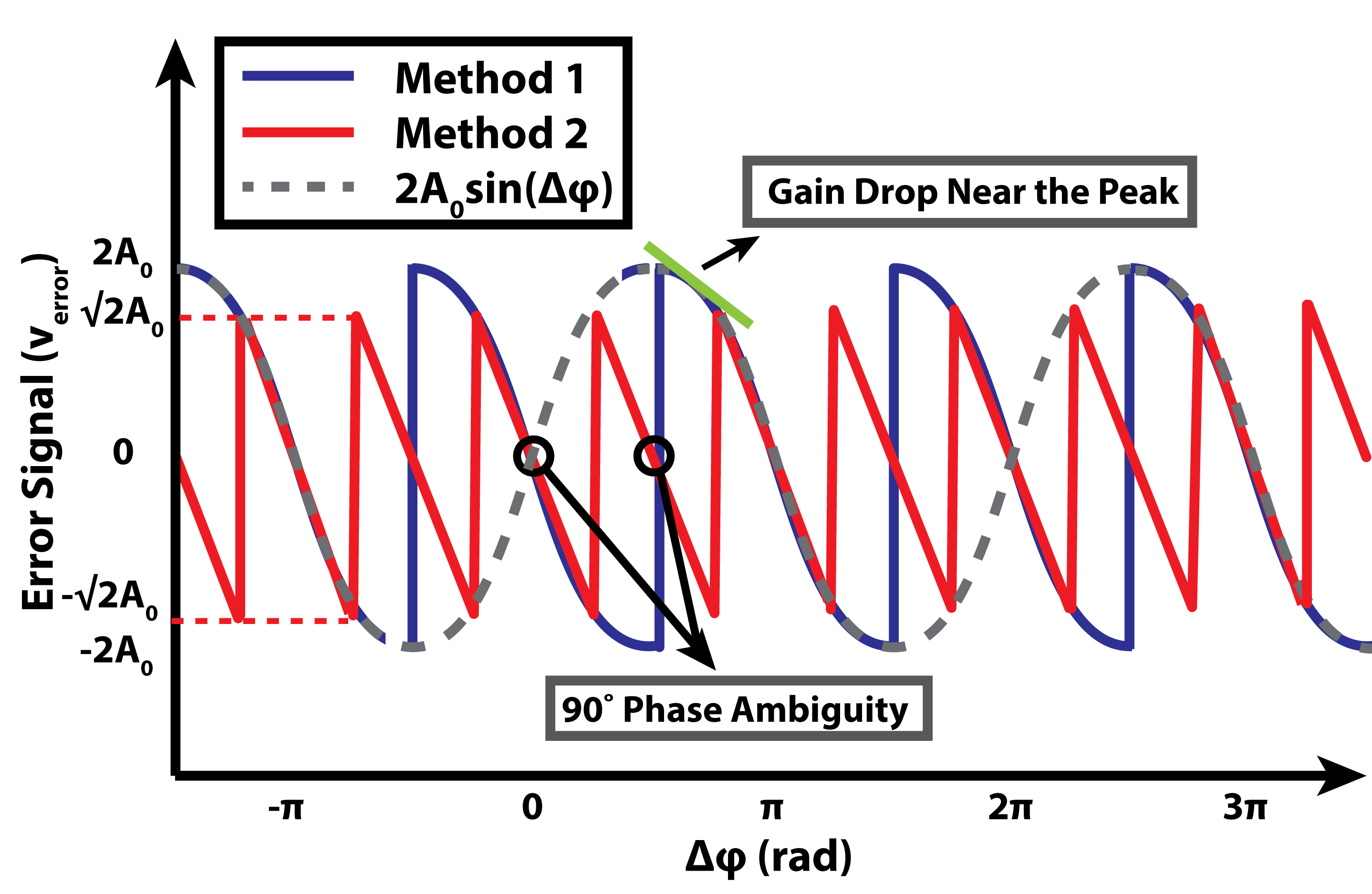}}
\caption{The error signal generated by Method 1 and Method 2.}
\label{Error Signal}
\end{figure}

Optical coherent communication relies on precise frequency and phase locking between the received signal and the local oscillator. Any phase error ($\Delta\phi$) between these paths induces I/Q cross-talk, severely impacting symbol/bit error rates (SER and BER)~\cite{I/Q_imbalance}. Fig.~\ref{Constellation} presents the constellation diagrams for QPSK/4-QAM and 4-offset-QAM, highlighting the effects of phase offset between the Rx and LO signals. In QPSK/4-QAM, a $\Delta\phi$ phase offset causes the entire constellation to rotate, which can be corrected using the Costas loop technique~\cite{ASP_USB_ODLL,ASP_USB_OPLL,ASP_IIT_PLL}. However, in offset-QAM, phase offset rotates the constellation center and distorts amplitudes with respect to the origin making the Costas loop approach ineffective~\cite{marziyeh_mwscas}. 
We have proposed a novel technique to detect the phase offset between the Rx and LO paths based on the average power or voltage difference between the in-phase (I) and quadrature (Q) signals in offset-QAM modulation.
Fig.~\ref{Methods}(a) illustrates the CPR block diagram for laser-forwarded links using a tunable phase shifter in the LO path. In this approach, the Tx laser output is split, with one portion modulated with data and the other transmitted to the Rx as the LO signal. The phase error between the Rx and LO signals is detected in the CPR block and then compensated using a phase shifter in the LO path. Two techniques for phase error detection are shown in Fig.~\ref{Methods}(b) and 2(c). The I and Q signals, before phase error compensation, are modeled by the following equations:
\begin{alignat}{1}
    I'(t) = (I(t) + A_0)cos(\Delta\phi) + (Q(t) + A_0)sin(\Delta\phi)\nonumber\\Q'(t) = (Q(t) + A_0)cos(\Delta\phi) - (I(t) + A_0)sin(\Delta\phi)
    \label{eq:I and Q}
\end{alignat}
where $A_0$ is a constant due to offset-QAM modulation and $I(t)$ and $Q(t)$ are $\pm \frac{A_{OMA}}{2}$. Then high-frequency portions of $I'(t)$ and $Q'(t)$ are filtered using low-pass filters, retaining the average voltage as shown below:
\begin{alignat}{1}
    I_{avg}(\Delta\phi) = A_0cos(\Delta\phi) + A_0sin(\Delta\phi)\nonumber\\
    Q_{avg}(\Delta\phi) = A_0cos(\Delta\phi) - A_0sin(\Delta\phi)
\end{alignat} 
In Method 1, the outputs of the low-pass filters in the I and Q paths are subtracted, generating an error signal equal to $2A_0 \sin{(\Delta\phi)}$. However, if the error signal is in the region where the slope is positive, the loop becomes unstable. To maintain stability, the summation of $I_{avg}(\Delta\phi)$ and $Q_{avg}(\Delta\phi)$ ($2A_0\cos{(\Delta\phi)}$) is used as a select signal to choose between $2A_0\sin{(\Delta\phi)}$ and -$2A_0\sin{(\Delta\phi)}$ in different regions. For example, when $\frac{-\pi}{2} < \Delta\phi < \frac{\pi}{2}$, and the error signal's slope is positive, $\cos{(\Delta\phi)}$ is positive and chooses $-2A_0\sin{(\Delta\phi)}$, and vice versa. This ensures that the error signal matches what is depicted in Fig.~\ref{Error Signal}.

In Method 2, $I_{avg}(\Delta\phi)$ and $Q_{avg}(\Delta\phi)$ are sent to limiting amplifiers (SGN) and mixers, then subtracted, similar to the Costas loop technique, producing an error signal shown in Fig.~\ref{Error Signal}~\cite{ASP_USB_ODLL}. This method generates a sawtooth error signal that maintains constant gain over the entire range. However, the periodicity of $\frac{\pi}{2}$ introduces phase ambiguity between the I and Q signals at the receiver\cite{ashok2021analog}. To resolve this, differential coding during modulation and demodulation is required.

While Method 1 avoids the $\frac{\pi}{2}$ phase ambiguity, its gain diminishes near the peaks, affecting the loop bandwidth and phase error compensation time. In this work, we adopt the phase detection technique based on Method 1.

\section{CPR Loop Dynamics}
\label{Loop Dynamics}
\begin{figure}
    \centering
    \includegraphics[width=\linewidth]{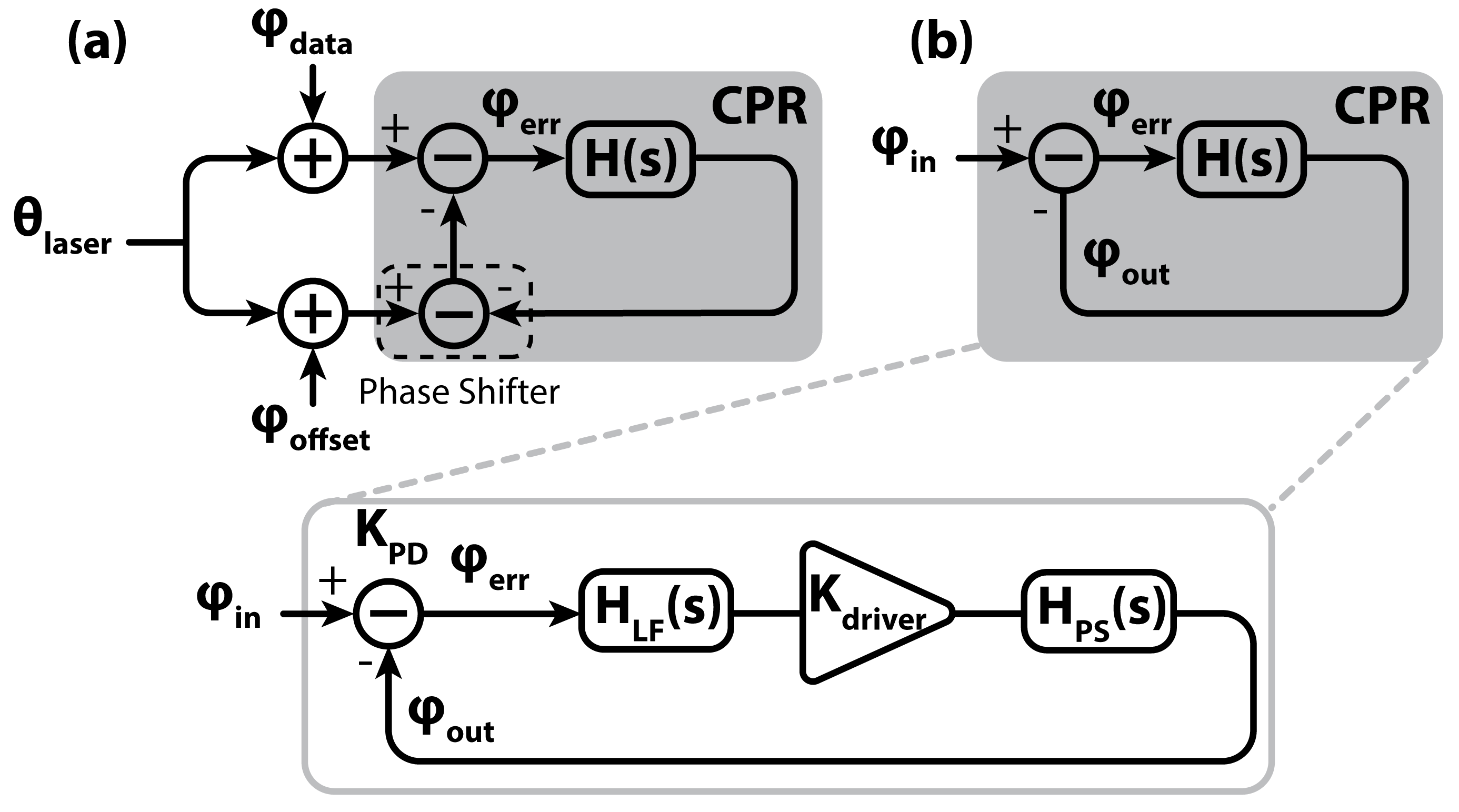}
    \caption{(a) The linear small-signal phase modeling of the CPR system (b) the simplified phase model after averaging.}
    \label{linear_model}
\end{figure}

In this section, the CPR loop dynamics are analyzed in terms of stability, bandwidth, residual phase error, and phase noise performance. 

In the transmitter, the electric field of the laser source can be expressed as:
\begin{equation}
E_{laser}(t) = E_0 e^{j(\omega_c t + \phi_{0,laser} + \phi_{pn}(t))}
\end{equation}
where \(E_0\), \(\omega_c\), \(\phi_{0,laser}\), and \(\phi_{pn}(t)\) denote the laser's electric field magnitude, center frequency, initial phase, and phase noise, respectively. This signal is subsequently modulated with offset-QAM data and transmitted over fiber. At the receiver, the Rx signal ($E_{Rx}$) beats with the LO signal which is a portion of the Tx laser forwarded to the receiver and generates a beat signal proportional to \(E_{Rx}(t)E_{LO}^*(t)\). Assuming a length mismatch of \(\Delta L\) between the Rx and LO paths, the beat signal at the receiver has a phase noise expressed as $\phi_{pn}(t) - \phi_{pn}(t - \tau_{\Delta L})$; where \(\tau_{\Delta L} = \frac{n \Delta L}{c}\), with \(n\) being the refractive index and \(c\) the speed of light~\cite{JLT_Phase_noise_16-QAM}. 
Fig.~\ref{linear_model}(a) illustrates the receiver's phase linear model, where \(\theta_{laser}(t) = \omega_c t + \phi_{laser,0} + \phi_{pn}(t)\). In this context, \(\phi_{data}(t)\) denotes the modulation phase, \(\phi_{offset}\) refers to the phase offset caused by the length mismatch between the Rx and LO paths, \(\phi_{err}\) is the phase error detected by the CPR loop, and \(H(s)\) represents the open-loop transfer function of the CPR. 

Since \(\phi_{data}(t)\) is initially filtered by low-pass filters, the loop model simplifies to the form shown in Fig.~\ref{linear_model}(b), where \(\phi_{in}(t) = \phi_{offset} + \phi_{pn}(t) - \phi_{pn}(t - \tau_{\Delta L})\). The linearized open-loop transfer function is expressed as follows:
\begin{equation}
    H(s) = K_{PD} K_{{driver}} \times H_{LF}(s) \times H_{PS}(s)
\end{equation}
\begin{equation*}
    H_{LF}(s) = \frac{K_{{LF}}(1 + \frac{s}{\omega_{{LF,z}}})}{1 + \frac{s}{\omega_{{LF,p}}}}
\end{equation*}
\begin{equation*}
    H_{PS}(s) = \frac{K_{{PS}}}{1 + \frac{s}{\omega_{{PS}}}}
\end{equation*}
where $K_{PD}$ is the phase error detector gain expressed as \(\frac{2\sqrt{2}}{\pi} I_{0}K_v\), with \(I_{0}\) indicating the average photocurrent of the I/Q signals, defined as \(I_{0} = 4|\overline{E_{I/Q}}| \times |E_{0,LO}| \times R_{PD}\). Here, \(|E_{0,LO}|\) and \(|\overline{E_{I/Q}}|\) correspond to the magnitudes of the electric field for the LO and the average I and Q signals, respectively. For DC-balanced I/Q data streams, \(|\overline{E_{I/Q}}|\) equals \(A_0\) as shown in Fig.~\ref{Constellation}. The photodiode responsivity is represented by \(R_{PD}\) and $K_v$ is the current to voltage conversion gain. It is important to note that in Method 1, \(K_{PD}\) is not constant and exhibits variations near the peak; however, for simplicity, it is approximated by its value in the linear region. \(K_{driver}\) refers to the gain of the phase shifter driver and \(K_{LF}\), \(\omega_{LF,z}\), and \(\omega_{LF,p}\) represent the gain, zero, and pole of the loop filter, respectively, while \(K_{PS}\) and \(\omega_{PS}\) denote the gain and electrical 3-dB bandwidth of the phase shifter.
\begin{table}[!t]
\caption{CPR loop parameters.}
\label{table parameters}
\centering
\begin{tabular}{|c|c|c|}
\hline
\textbf{$\boldsymbol{K_{PD}}$} & \textbf{$\boldsymbol{K_{LF}}$} & \textbf{$\boldsymbol{K_{driver}}$} \\ \hline
2.55e-2 (V/rad) & 1.2e3 (V/V) & 2 (V/V) \\ \hline
\textbf{$\boldsymbol{K_{PS}}$} & \textbf{$\boldsymbol{f_{LF,z}}$} & \textbf{$\boldsymbol{f_{LF,p}}$} \\ \hline
15.7 (rad/V) & 0.8MHz & 6kHz \\ \hline
\multicolumn{3}{|c|}{\textbf{$\boldsymbol{f_{PS}}$}} \\ \hline
\multicolumn{3}{|c|}{2kHz} \\ \hline
\end{tabular}
\end{table}

\begin{figure}
    \centering
    \includegraphics[width=\linewidth]{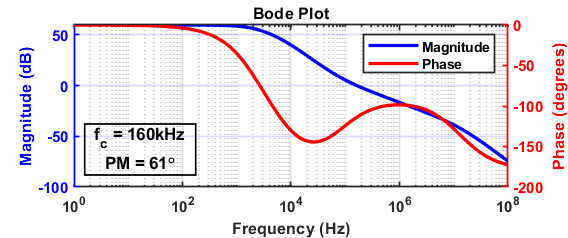}
    \caption{Magnitude and phase plots of the open-loop transfer function of the CPR system.}
    \label{bode plot}
\end{figure}

Fig. ~\ref{bode plot} presents the Bode plot of the proposed CPR system, with parameters specified in Table~\ref{table parameters}. The system demonstrates a phase margin of $61^\circ$, a crossover frequency of 160 kHz, and a closed-loop bandwidth of 216 kHz.
\subsection{Residual Phase Error}
The residual phase error can be determined from the phase error transfer function, given by $\frac{\phi_{err}(s)}{\theta_{in}(s)} = \frac{1}{1 + H(s)}$. The static phase error, resulting from a constant input phase offset of $\phi_0$, can be calculated from:
\begin{equation}
    \begin{aligned}
    \lim_{t \to \infty} \phi_{err}(t) &= \lim_{s \to 0} s \phi_{err}(s) \\
    &= \lim_{s \to 0} s \times \frac{\phi_0/s}{1 + H(s)} \\
    &= \frac{\phi_0}{1+H(0)} \\
    &= \frac{\phi_0}{1+K_{PD} K_{LF} K_{driver} K_{PS}}
    \end{aligned}
\end{equation}

Where, for the parameters given in Table~\ref{table parameters}, the static phase error is equal to \(\frac{\phi_0}{960}\) for \(-\frac{\pi}{4} \leq \phi_0 \leq \frac{\pi}{4}\) (the calculated loop gain is based on the linear approximation of \(K_{PD}\)). For instance, with \(\phi_0 = \frac{\pi}{4}\), the resulting phase error is approximately \(0.8 \, \text{mrad}\). The residual phase error can be further minimized by increasing the loop gain.

\subsection{Laser Phase Noise}
The phase noise of a Lorentzian-shaped laser linewidth can be described as a random walk Wiener process with a power spectral density (PSD) of \(S_{\phi_{pn}}(f) = \frac{\Delta\nu}{2\pi f^2}\)~\cite{JLT_Phase_noise_16-QAM}, where \(\Delta\nu\) is the laser linewidth. 
To accurately calculate the variance of the resulting Gaussian phase noise, we need to take into account the effect of the CPR loop. Taking the Fourier transform of the detected signal phase noise, we obtain:
\begin{equation}
\theta_{pn}(f) = \phi_{pn}(f) (1 - e^{-j2\pi f\tau_{\Delta L}})\times\frac{1}{1+H(f)}
\end{equation}

where \(\theta_{pn}(f)\) and \(\phi_{pn}(f)\) are the phase noises of the beating signal and original laser in the frequency domain, and \(H(f)\) is the CPR open-loop transfer function as mentioned before. Therefore, the power spectral densities are related by:
\begin{equation}
\label{Eq:PN PSD calculation}
\begin{aligned}
    S_{\theta_{pn}}(f) &= S_{\phi_{pn}}(f)\times\frac{|1 - e^{-j2\pi f\tau_{\Delta L}}|^2}{|1+H(f)|^2}\\
    &= 2S_{\phi_{pn}}(f)\times\frac{1-\cos{2\pi f\tau_{\Delta L}}}{|1+H(f)|^2}
\end{aligned}
\end{equation}

The total phase noise power can be calculated from the following equation:
\begin{equation}
    \begin{aligned}
        \phi_{pn,tot}^2 &= \int_{-\infty}^{\infty} S_{\theta_{pn}}(f)\,df  
    \end{aligned}
\end{equation}

\begin{figure}
    \centering
    \includegraphics[width=\linewidth]{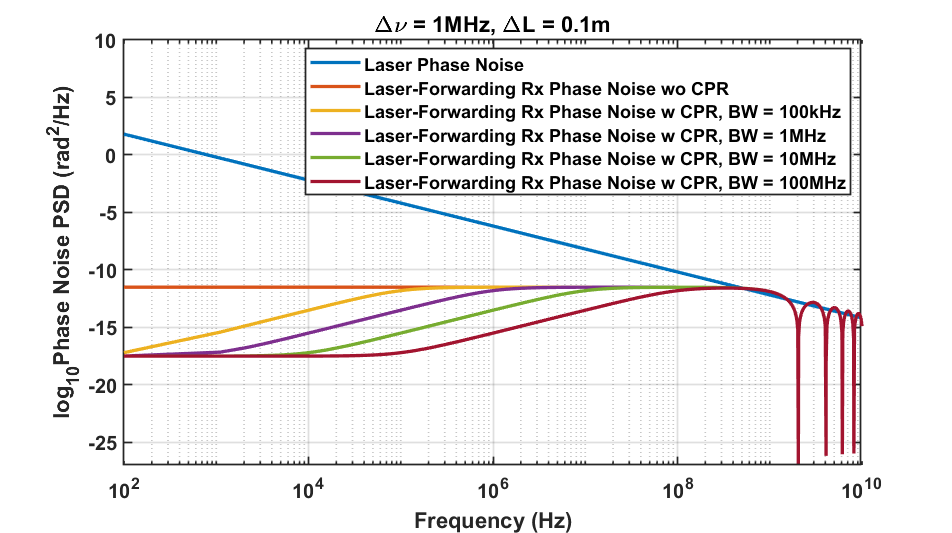}
    \caption{Effect of laser forwarding technique and CPR loop bandwidth on the PSD of the beat signal phase noise ($\Delta \nu$ and $\Delta L$ denote laser line-width and forwarding path mismatch length, respectively).}
    \label{fig:PSD vs BW}
\end{figure}

Fig.~\ref{fig:PSD vs BW} shows the PSD of the beat signal’s phase noise for different CPR loop bandwidths. In this figure, the laser-forwarding technique shapes the laser phase noise depending on $\Delta L$ as shown in Eq.~\ref{Eq:PN PSD calculation}. The term $\frac{1}{|1+H(f)|^2}$ exhibits high-pass behavior so the CPR loop bandwidth further filters the phase noise up to its cutoff. Increasing the CPR loop bandwidth reduces the phase noise power more effectively.
As mentioned before, the phase noise of the beat signal can be represented as Gaussian noise with a mean of 0 and variance of \(\phi_{pn,tot}^2\). With this, we can calculate the theoretical BER with respect to the signal-to-noise ratio (SNR) and laser phase noise.

\subsubsection{Impact of Laser Phase Noise on BER in 4-Offset-QAM} 
\begin{figure}
    \centering
    \includegraphics[width=\linewidth]{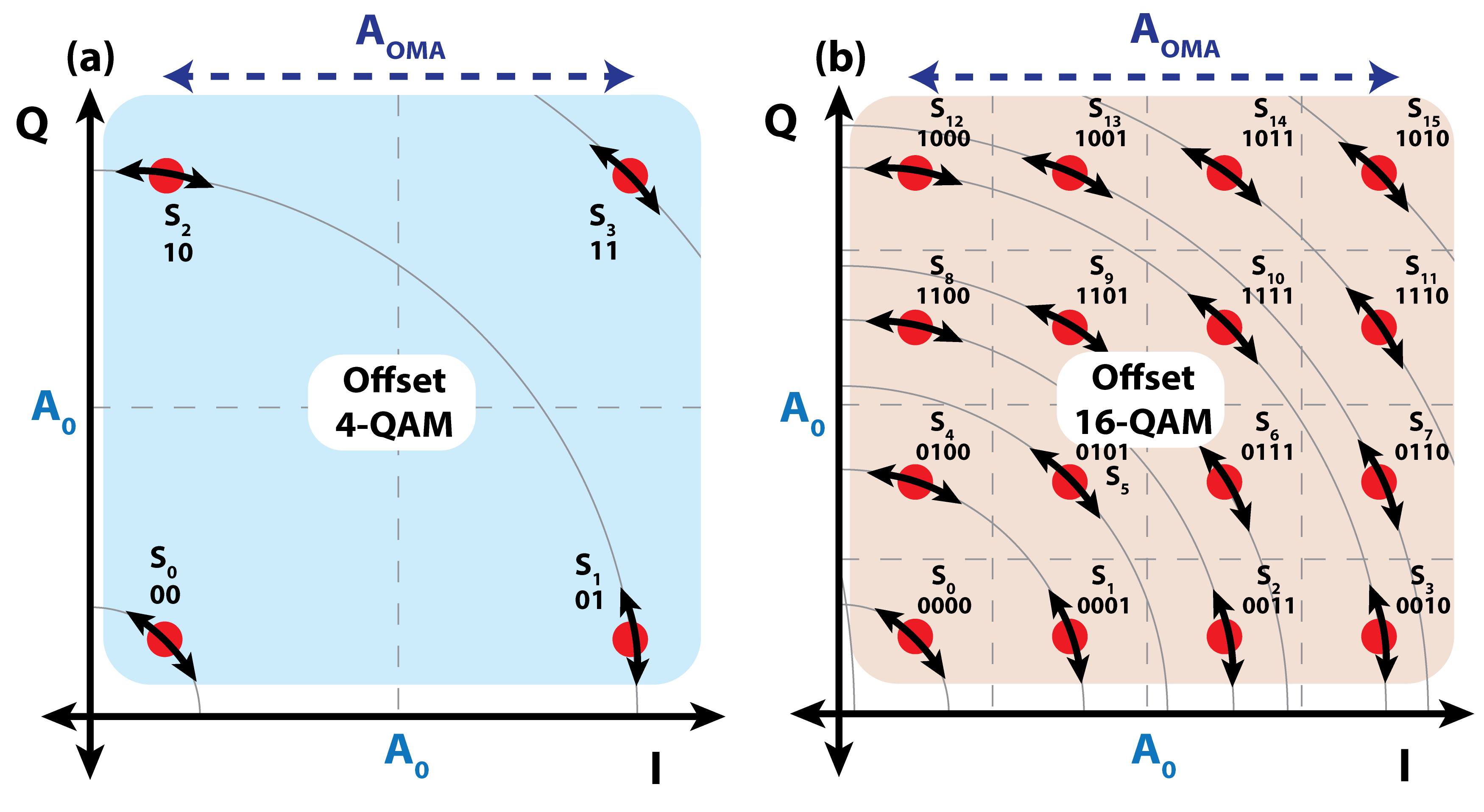}
    \caption{Effect of laser phase noise on (a) 4-offset-QAM and (b) 16-offset-QAM constellations.}
    \label{Phase Noise Effect}
\end{figure}

Fig.~\ref{Phase Noise Effect}(a) illustrates the constellation diagram of 4-Offset-QAM, where symbols rotate on circles with radius proportional to their symbol magnitude. The symbol error rate and bit error rate are derived as follows~\cite{JLT_Phase_noise_16-QAM}:
\begin{equation}
\label{Eq:BER/SER}
\begin{aligned}
        SER &= \sum_{i=0}^{n-1} P(S_i) \times P(e|S_i) \\
        BER &= \frac{SER}{log_2(n)}
\end{aligned}
\end{equation}

where $n$ is the modulation depth (equal to 2 for 4-offset-QAM), \( P(S_i) \) is the probability of symbol \( i \) occurring (equal to 0.25 for 4-offset-QAM), and \( P(e|S_i) \) is the conditional probability that the transmitted symbol \( i \) is incorrectly detected as one of the other symbols by the receiver. \( P(e|S_i) \) occurs when either the I or Q terms, or both, are incorrectly detected and can be calculated by~\cite{JLT_Phase_noise_16-QAM}:
\begin{equation}
\label{Eq:P_error equation}
       P(e|S_i) = P(e_I|S_i) + P(e_Q|S_i) - P(e_I|S_i)P(e_Q|S_i)
\end{equation}

Offset-QAM exhibits a different BER expression compared to QPSK, attributed to its offset value. Recalling from Eq.~\ref{eq:I and Q}, we can simplify I and Q equations to the following format:
\begin{equation}
\label{Eq:I/Q}
    \begin{aligned}
            I'(t) &= \sqrt{I^2(t)+Q^2(t)}\cos{(\theta_{pn}(t) - \arctan{(\frac{Q(t)}{I(t)})})}\\
            &+ \sqrt{2}A_0\cos{(\theta_{pn}(t) - \frac{\pi}{4})}\\
            Q'(t) &= \sqrt{I^2(t)+Q^2(t)}\cos{(\theta_{pn}(t) + \arctan{(\frac{I(t)}{Q(t)})})}\\
            &+ \sqrt{2}A_0\cos{(\theta_{pn}(t) + \frac{\pi}{4})}
    \end{aligned} 
\end{equation}

Where $A_0$ is the offset-QAM offset, $I(t)$ and $Q(t)$ are $\pm \frac{A_{OMA}}{2}$, and $\theta_{pn}(t)$ is the beat signal phase noise. Next, we show the conditional error probability for the \( S_1 \) symbol, and the same calculation method applies to the remaining symbols. 
The error in detecting the \( S_1 \) symbol occurs when the combined effects of noise and phase noise causes \( I_{S_1} \) to fall below \( A_0 \) and/or \( Q_{S_1} \) to exceed \( A_0 \). The conditional error probability for the \( S_1 \) symbol is given below:
\begin{equation}
    \begin{aligned}
        P_{e_I|S_1}(\theta_{pn}) &= P(I_{S_1} + n(t) < A_0) \\
        &=P(n(t) < -\frac{\sqrt{2}}{2}A_{OMA}\cos{(\theta_{pn}+\frac{\pi}{4})}\\ 
        &\hspace{42pt}  + A_0(1-\sqrt{2}\cos{(\theta_{pn} - \frac{\pi}{4})}))
    \end{aligned}
\end{equation}

where \( n(t) \) is the channel additive white Gaussian noise (AWGN) with a PSD of \( N_0\). As seen in this equation, depending on where the location of the center ($A_0$) is, the phase noise effect on BER varies. As the center moves closer to the origin, phase noise performance improves. Assuming $A_0 = mA_{OMA}$, we can simplify the aforementioned equation into:
\begin{equation}
    \begin{aligned}
        P_{e_I|S_1}(\theta_{pn}) &=P(n(t) < A_{OMA}(-\frac{\sqrt{2}}{2}\cos{(\theta_{pn}+\frac{\pi}{4})}\\ 
        &\hspace{84pt}  + m(1-\sqrt{2}\cos{(\theta_{pn} - \frac{\pi}{4})}))\\
        &=\frac{1}{2}erfc(-\frac{A_{OMA}}{\sqrt{N_0}}(m-(m+\frac{1}{2})\cos{\theta_{pn}}
        \\&\hspace{84pt}-(m-\frac{1}{2})\sin{\theta_{pn}}))
    \end{aligned}
\end{equation}

Where erfc(.) is the complementary error function defined as:
\begin{equation}
    \operatorname{erfc}(x) = \frac{2}{\sqrt{\pi}} \int_x^{+\infty} e^{-z^2} \, dz
\end{equation}

The total conditional error probability due to the phase noise can be calculated from~\cite{JLT_Phase_noise_16-QAM}:
\begin{equation}
        P(e_I|S_1) = \int_{-\pi}^{\pi} P_{e_I|S_1}(\theta_{pn})\times f_{pn}(\theta_{pn})\,d\theta_{pn}
\end{equation}

where \( f_{pn}(\theta_{pn}) \) is the PDF of the beat signal Gaussian phase noise \( \left(\frac{1}{\sqrt{2 \pi \sigma^2}} e^{-\frac{\theta_{pn}^2}{2 \sigma^2}}\right) \) with the variance calculated above (\( \sigma^2=\phi_{pn,tot}^2 \)). Similarly, we can calculate \( P(e_Q|S_1) \).
\begin{figure*}
\centering
\subfloat[]{\includegraphics[width=\linewidth]{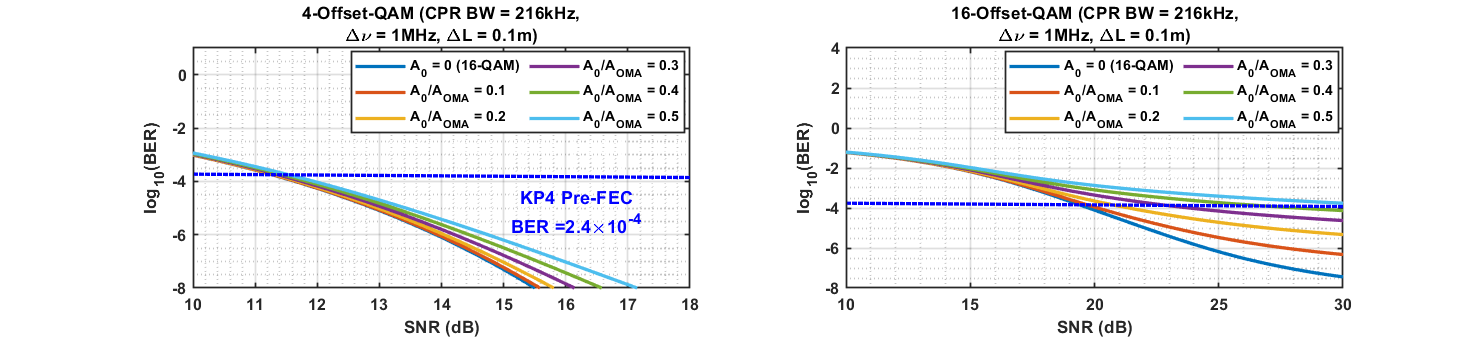}}
\hfill
\subfloat[]{\includegraphics[width=\linewidth]{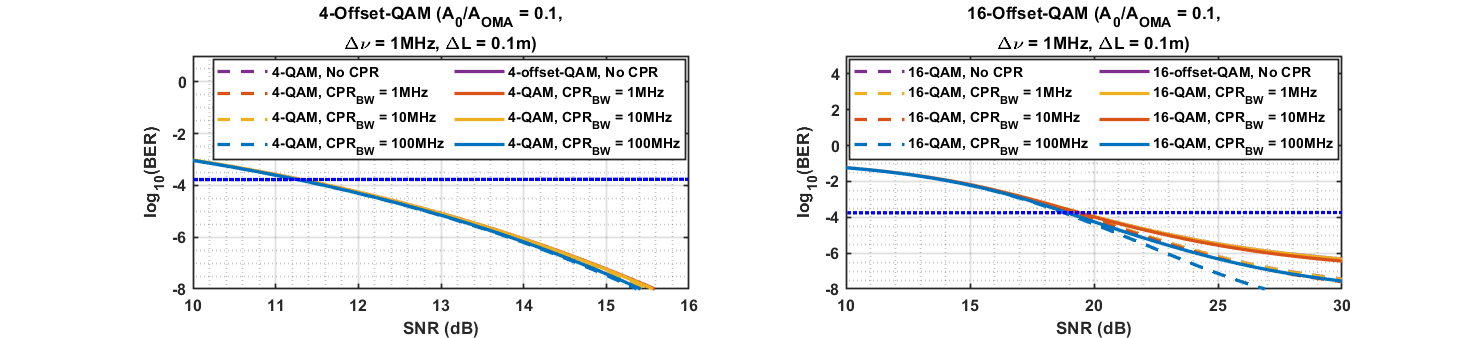}}
\hfill
\subfloat[]{\includegraphics[width=\linewidth]{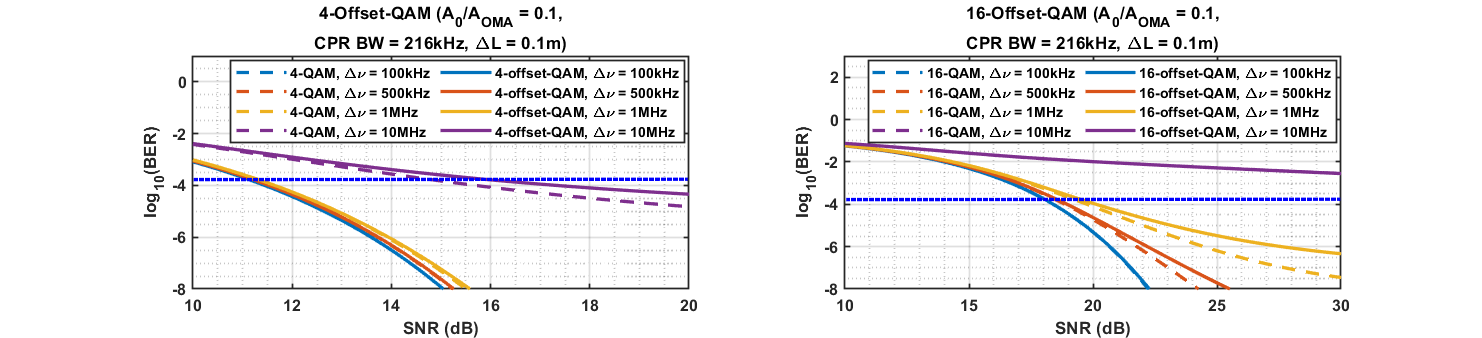}}
\hfill
\subfloat[]{\includegraphics[width=\linewidth]{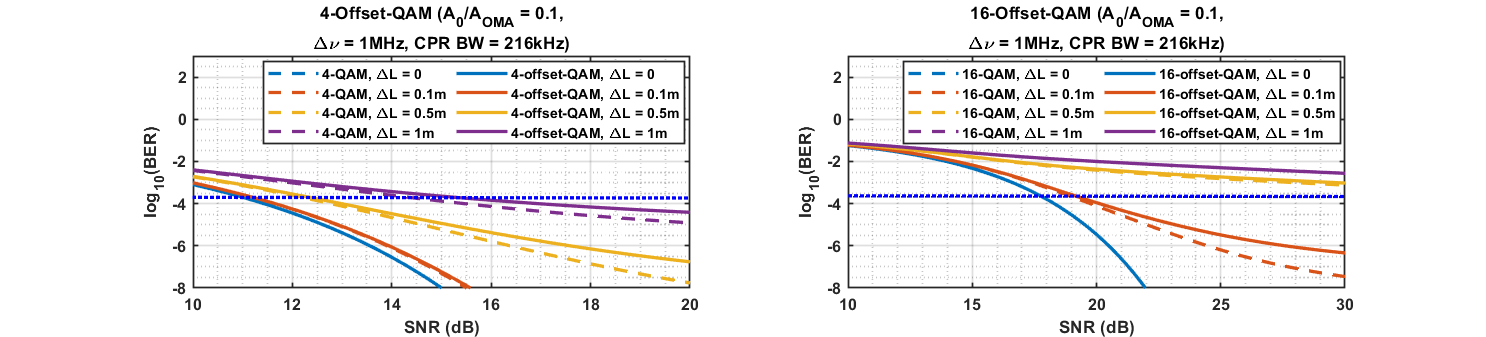}}
\caption{Theoretical BER vs. SNR for 4- and 16-Offset-QAM for various (a) constellation offset (\( A_0 \)) values, (b) CPR loop bandwidths, (c) laser linewidths, and (d) LO/RX fiber length mismatch. Dashed lines indicate the FEC limits of \( 2.4\times10^{-4} \) for KP4 FEC.}
\label{fig:BER}
\end{figure*}
With \( P(e_Q|S_1) \) and \( P(e_I|S_1) \) calculated, we can determine \( P(e|S_1) \) and subsequently the error probability of the other symbols, as well as the SER and BER from Eq.~\ref{Eq:BER/SER}. To have a universal comparison of BER versus SNR, the SNR is defined as the ratio of the average energy per symbol to the noise power spectral density, denoted by \( \frac{E_s}{N_0} \). The average symbol energy \( E_s \) can be calculated from:
\begin{equation}
\label{Eq:E_avg}
    E_s = \frac{1}{log_2(m)}\sum_{i=0}^{m-1} (I'(t)_{|\Delta\phi=0}-A_0)^2 + (Q'(t)_{|\Delta\phi=0}-A_0)^2
\end{equation}

 For 4-offset-QAM, we have $E_s = \frac{A_{OMA}^2}{2}$. To investigate the impact of the offset-QAM center location (\( A_0 \)), we calculated the BER versus SNR for various values of \( A_0 \). To evaluate BER performance, we have compared all results with a KP4 forward error correction (FEC) standard which requires a BER threshold of \( 2.4 \times 10^{-4} \)~\cite{KP4_FEC}. In practice, any other FEC standard can be chosen depending on the link latency and application-level specifications. The results shown in Fig.~\ref{fig:BER}(a) demonstrate that for small offsets, the BER performance of regular 4-QAM and 4-offset-QAM is nearly identical.

Next, we examine the effect of the CPR loop bandwidth on the BER by calculating the BER versus the SNR for different loop bandwidth values for \( \frac{A_0}{A_{OMA}} = 0.1 \). As depicted in Fig.~\ref{fig:BER}(b), the effect of loop bandwidth on BER for 4-offset-QAM is negligible. 

We also studied the impact of laser linewidth on BER. DFB lasers typically exhibit linewidths in the range of 1-10 MHz \cite{JLT_Phase_noise_16-QAM}. As shown in Fig.~\ref{fig:BER}(c), as the linewidth increases from 100kHz to 1MHz, the required SNR penalty to achieve a BER of \( 2.4 \times 10^{-4} \) is approximately 0.2dB.

Furthermore, we investigated the effect of length mismatch on BER. As illustrated in Fig.~\ref{fig:BER}(d), for a laser linewidth of 1MHz, when the length mismatch exceeds 50cm, the BER drops significantly. This highlights the importance of carefully managing the splitting of the laser output between the transmitter and local oscillator LO to ensure that the length mismatch remains within a few centimeters.

\subsubsection{Effect of Laser Phase Noise on BER for 16-Offset-QAM}
Next, we investigate the effect of laser phase noise on BER for 16-offset-QAM modulation, using a procedure analogous to that for 4-offset-QAM. Fig.~\ref{Phase Noise Effect}(b) presents the constellation diagram of 16-Offset-QAM, illustrating the symbol shifts caused by phase noise. The values of I and Q fall within the set $\lbrace \pm\frac{A_{OMA}}{6}, \pm\frac{3A_{OMA}}{6} \rbrace$, with the center located at ($A_0$, $A_0$). The SER can be calculated using Eq.~\ref{Eq:BER/SER}, setting $n$ to 16 and $P(S_i)$ to 0.0625. The conditional error probability of the symbols can then be calculated from Eq.~\ref{Eq:P_error equation}. Here, we demonstrate the calculation of the conditional error probability for the $S_1$ symbol ($-\frac{A_{OMA}}{6}+A_0$, $-\frac{3A_{OMA}}{6}+A_0$).
The error in detecting the \( S_1 \) symbol occurs when the combined effects of noise and phase noise cause \( I_{S_1} \) to fall below \( A_0-\frac{A_{OMA}}{3} \) or exceed \(A_0\), and/or \( Q_{S_1} \) to exceed \( A_0-\frac{A_{OMA}}{3} \). Assuming $A_0 = mA_{OMA}$, the conditional error probability for the \( S_1 \) symbol can be calculated as follows:
\begin{equation}
    \begin{aligned}
        P_{e_I|S_1}(\theta_{pn}) &= P(I_{S_1} + n(t) < A_0-\frac{A_{OMA}}{3})\\ 
        &\hspace{10pt}+P(I_{S_1} + n(t) > A_0) \\
        &=\frac{1}{2}erfc({-\frac{A_{OMA}}{\sqrt{N_0}}}((m-\frac{1}{3})-(m-\frac{1}{6})\cos{\theta_{pn}}
        \\&\hspace{85pt}-(m-\frac{3}{6})\sin{\theta_{pn}}))\\
        &+\frac{1}{2}erfc({\frac{A_{OMA}}{\sqrt{N_0}}}(m-(m-\frac{1}{6})\cos{\theta_{pn}}
        \\&\hspace{75pt}-(m-\frac{3}{6})\sin{\theta_{pn}}))
    \end{aligned}
\end{equation}
\begin{equation}
    \begin{aligned}
        P_{e_Q|S_1}(\theta_{pn}) &= P(Q_{S_1} + n(t) > A_0-\frac{A_{OMA}}{3})\\ 
        &=\frac{1}{2}erfc({\frac{A_{OMA}}{\sqrt{N_0}}}((m-\frac{1}{3})-(m-\frac{3}{6})\cos{\theta_{pn}}
        \\&\hspace{85pt}+(m-\frac{1}{6})\sin{\theta_{pn}}))\\
    \end{aligned}
\end{equation}
Similarly, we can determine the conditional error probability for the remaining symbols to calculate the total SER, and therefore, the BER. From Eq.~\ref{Eq:E_avg}, we have $E_s = \frac{5A_{OMA}^2}{18}$. The results shown in Fig.~\ref{fig:BER}(a) indicate that for a small offset (\( \frac{A_0}{A_{OMA}} = 0.1 \)), the BER performance of regular 16-QAM and 16-offset-QAM are nearly identical. However, as the offset increases, such as when \( \frac{A_0}{A_{OMA}} = 0.5 \), significant SNR penalties are observed. To provide reasonable input range for the CPR system and minimize the SNR penalty for 16-offset-QAM, \( \frac{A_0}{A_{OMA}} = 0.1 \) is chosen.

As shown in Fig.~\ref{fig:BER}(b), for a BER of \( 2.4 \times 10^{-4} \) with a linewidth of 1MHz and a length mismatch of 10cm, a CPR bandwidth of 10MHz and below has about 1dB SNR penalty compared to the 100MHz bandwidth. As demonstrated in Fig.~\ref{fig:BER}(c), increasing the linewidth from 100kHz to 1MHz results in 2dB SNR penalty. Finally, as depicted in Fig.~\ref{fig:BER}(d), for a DFB laser with a linewidth of 1MHz, a length mismatch of 10cm results in SNR penalty of 1dB compared to the case with no length mismatch and, consequently, no phase noise. These findings highlight the critical importance of selecting the offset-QAM center location and DFB laser linewidth, as well as designing the system to minimize the length mismatch between the LO and Rx paths. It is important to note that we have control over the length mismatch, which can be adjusted by tuning the fiber length to significantly reduce the SNR penalty.

\section{Coherent Receiver Circuit Implementation}
\label{Circuit}
\subsection{Offset-QAM Transmitter Modeling}
The 4-offset-QAM transmitter was modeled using ideal spectre models supporting up to $100\text{GBaud}$ to primarily evaluate the performance of the proposed CPR system at high baud rates. Additionally, the transmitter was extended to support 16-offset-QAM to assess the effectiveness of the proposed CPR system for higher-order offset-QAM modulation formats.

\subsection{Circuit Implementation of Offset-QAM CPR System}
\begin{figure*}
\centerline{\includegraphics[width=\linewidth]{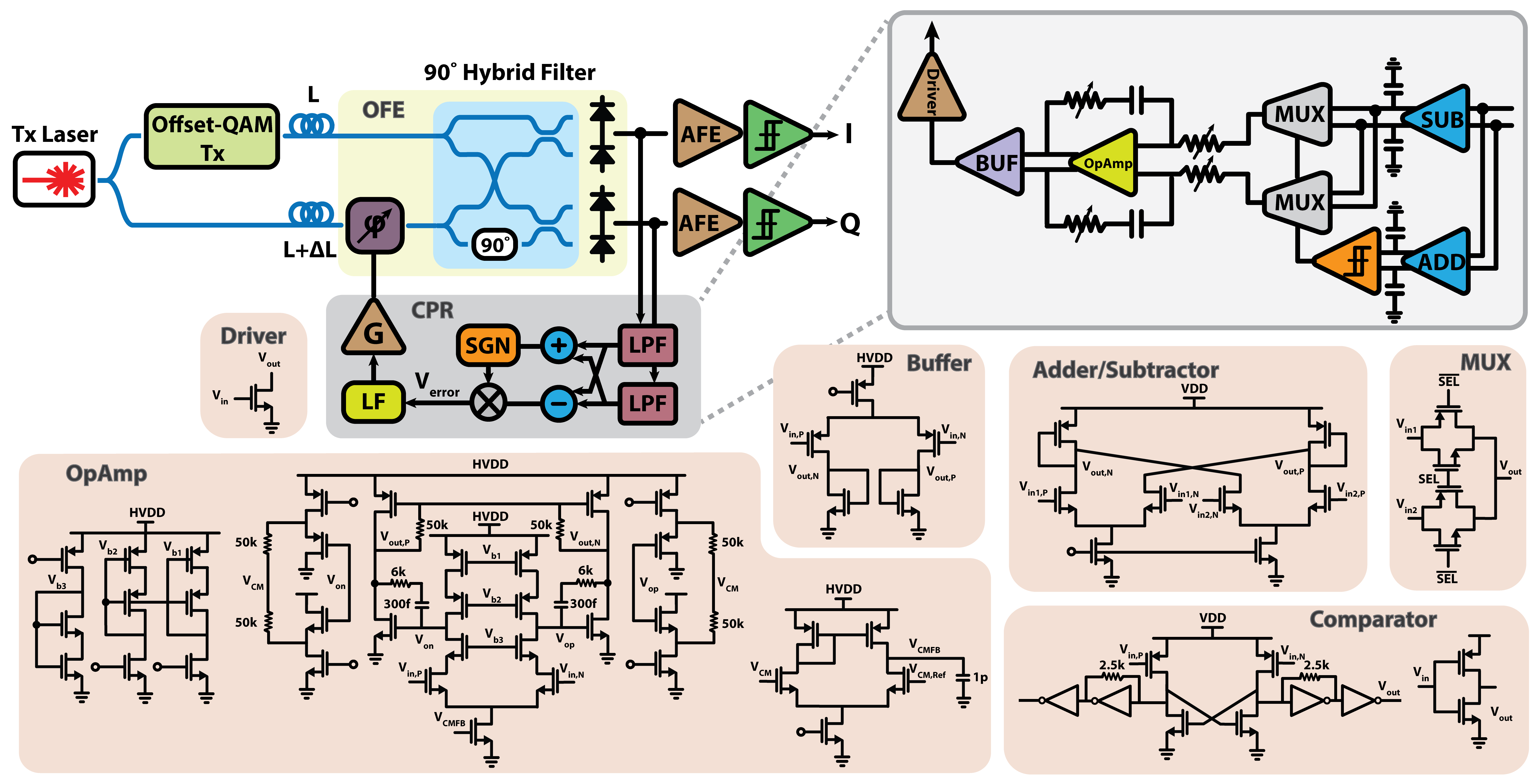}}
\caption{The circuit-level block diagram of the proposed Offset-QAM receiver.}
\label{Block diagram}
\end{figure*}
Fig.~\ref{Block diagram} shows the block diagram of the coherent receiver that can be used for any levels of offset-QAM modulation. A detailed circuit implementation of the coherent receiver is provided below.

\subsubsection{Optical Front-End (OFE)}
The local laser is forwarded from the transmitter side through a fiber. A thermal phase shifter with a 2kHz bandwidth in the LO path is used to adjust the LO phase via the CPR loop. At the receiver, the Rx and LO signals beat together in a $90^{\circ}$ optical hybrid filter, followed by the photodetectors (PDs) with 50GHz bandwidth that decode the differential I and Q signals. All optical components utilized in the OFE are actual PDK components, enabling accurate system-level performance evaluation.

\subsubsection{Carrier Phase Recovery System}
In the carrier phase recovery block, the I and Q differential outputs are differenced in current mode using a subtractor, followed by a $100 \, \text{fF}$ capacitor for initial high-speed data filtering, generating an error signal proportional to $\sin(\Delta\phi)$. The error signal is then processed by an adder and a comparator, which select only the negative slope to ensure loop stability. The comparator is implemented using a non-clocked strong-arm latch comparator followed by inverters, with the first stage setting the bias point of the comparator.

To extract the average of the I and Q difference, a low-pass filter with a very low cutoff frequency is required. This is achieved using an active loop filter with high DC gain to guarantee zero remaining phase error between the LO and Rx signals. The loop filter consists of a differential operational amplifier (OpAmp) with $62 \, \text{dB}$ gain and $2 \, \text{pF}$ capacitive feedback in series with a resistive-DAC, creating a variable zero that enhances both bandwidth and loop stability. The OpAmp architecture is shown in Fig.~\ref{Block diagram}. The first stage is a telescopic design with common-mode feedback, while the second stage is a common-source amplifier with $100 \, \text{k}\Omega$ resistive feedback. The driver is a single transistor that continuously adjusts the LO phase shifter’s phase until the loop locks and the phase offset is zero. The VDD and HVDD are 1V and 1.8V, respectively. $v_{error}$ provides a swing range of $250mV$ to $1.8V$ that can compensate for approximately $5\pi$ phase offset between the Rx and LO signals. 
\section{Simulation Results}
\label{sim results}
\begin{figure*}[htbp]
\centerline{\includegraphics[width=\linewidth]{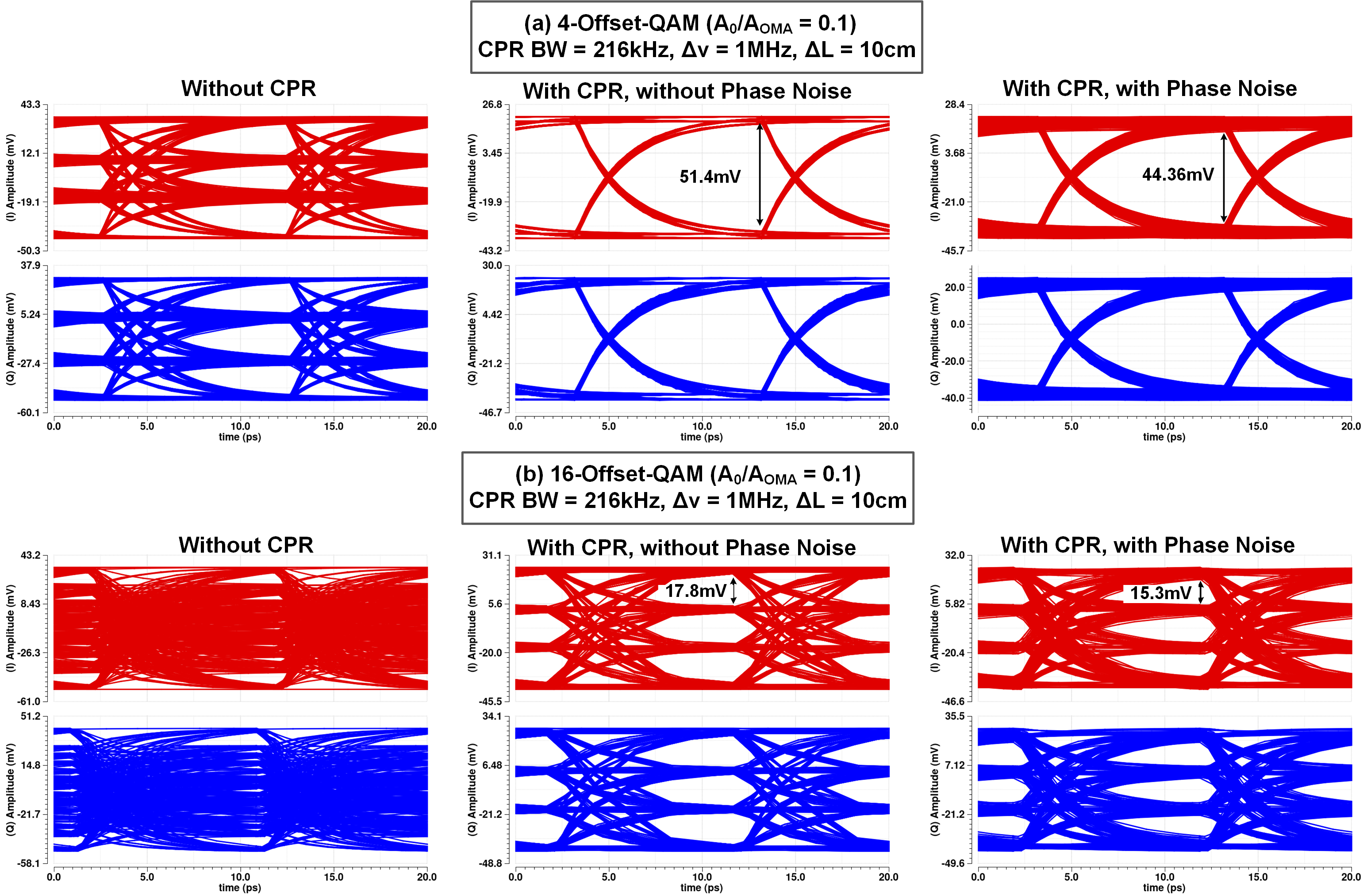}}
\caption{Simulated eye-diagrams of the I and Q channels at the AFE input nodes, for a $100GBaud$ offset-QAM receiver with an example phase offset of $\pi/4$ at: (a) $200Gb/s$ for 4-offset-QAM (b) $400Gb/s$ for 16-offset-QAM.}
\label{measurements}
\end{figure*}
We have designed and simulated the proposed laser-forwarded carrier phase recovery system using GlobalFoundry’s monolithic 45nm silicon photonics PDK models in Cadence. The parameter values used in this design are listed in Table~\ref{table parameters}. The baud rate is $100\text{GBaud}$, corresponding to data rates of $200\text{Gb/s}$ for 4-offset-QAM and $400\text{Gb/s}$ for 16-offset-QAM. Eye diagrams artifacts are mainly due to the bandwidth limitations of the photodetectors ($50 \, \text{GHz}$ with a $\sim 10\text{fF}$ capacitance) which can be improved using equalization techniques. The laser has a linewidth of $1\text{MHz}$, and there is a $0.1\text{m}$ length mismatch between the Rx and LO paths. The CPR loop bandwidth is $216\text{kHz}$. In these simulations, \( A_0 \) and \( A_{\text{OMA}} \) are set to \( 250 \,\mu\text{A} \) and \( 25 \,\mu\text{A} \), and the AFE first stage is modeled as a \( 220 \,\Omega \) resistor. 

To evaluate the functionality of the design, we performed closed-loop simulations for a phase offset of $\Delta \phi = \pi/4$ between the Rx and LO signals which is applied by an extra phase shifter in the LO path. Fig.~\ref{measurements}(a) shows the eye diagram of I and Q signals for 4-offset-QAM at the PDs output (input of the AFE) under three different scenarios: without the CPR loop, with the CPR loop and no phase noise, and with the CPR loop in the presence of phase noise. As can be seen in this figure, the CPR loop is able to cancel the phase offset with less than $1^\circ$ residual phase error. However, in the presence of phase noise, the eye opening is reduced. 

Fig.~\ref{measurements}(b) presents the results for $\Delta \phi = \pi/4$ for 16-offset-QAM. As seen in the figure, the proposed architecture successfully cancels the phase offset for 16-offset-QAM without requiring any changes to the design.

In the presence of phase noise, to achieve the same performance as in the absence of phase noise, we must reduce the phase noise power. This can be achieved by using a laser with a narrower linewidth, minimizing the length mismatch between the Rx and LO paths, or increasing the effective CPR closed-loop bandwidth, which is mainly limited by the thermal phase shifter bandwidth in this process.

\section{Conclusion}
\label{conclusion}
We have proposed, modeled, and simulated a DSP-free offset-QAM coherent receiver architecture that can perform CPR for any offset-QAM modulation level (e.g., 4, 16, 64) without architectural adjustments that can solve the scalability problem of previously proposed DSP-free CPR systems for QAM modulation. Combined with laser-forwarding, we believe this method can become a solution for
future low power coherent links inside data centers. We used electro-optical spectre models in Cadence using the GF45SPCLO for simulation. Received eye-diagram operating at a baud rate of 100Gbaud at the PD outputs are measured for both 4- and 16-offset-QAM.  Additionally, we studied the effects of laser linewidths, LO/Rx fiber length mismatch, and CPR loop bandwidth on BER. We have also elaborated on the circuit block implementation of key loop components. 

Realistic link scenarios assuming 1MHz laser linewidth and $10cm$ fiber length mismatch show that a 400Gb/s 16-Offset-QAM link with an SNR of $19dB$ can meet the KP4 FEC requirements. Our proposed approach can become a solution to realize DSP-free linear coherent optical transceivers (both pluggables and CPO) for 400Gb/s links and beyond. 



\bibliographystyle{IEEEtran}
\bibliography{reference}
\section*{Acknowledgment}
This work is supported by NSF CAREER (ECCS- 2142996), NSF ENG-SEMICON (ECCS-2430776) and SRC PKG Programs.  The authors would like to thank GlobalFoundries for providing silicon fabrication and PDK through the 45SPCLO university partnership program. 
\vspace{0pt}
\begin{IEEEbiography}[{\includegraphics[width=1in,height=1.25in,clip,keepaspectratio]{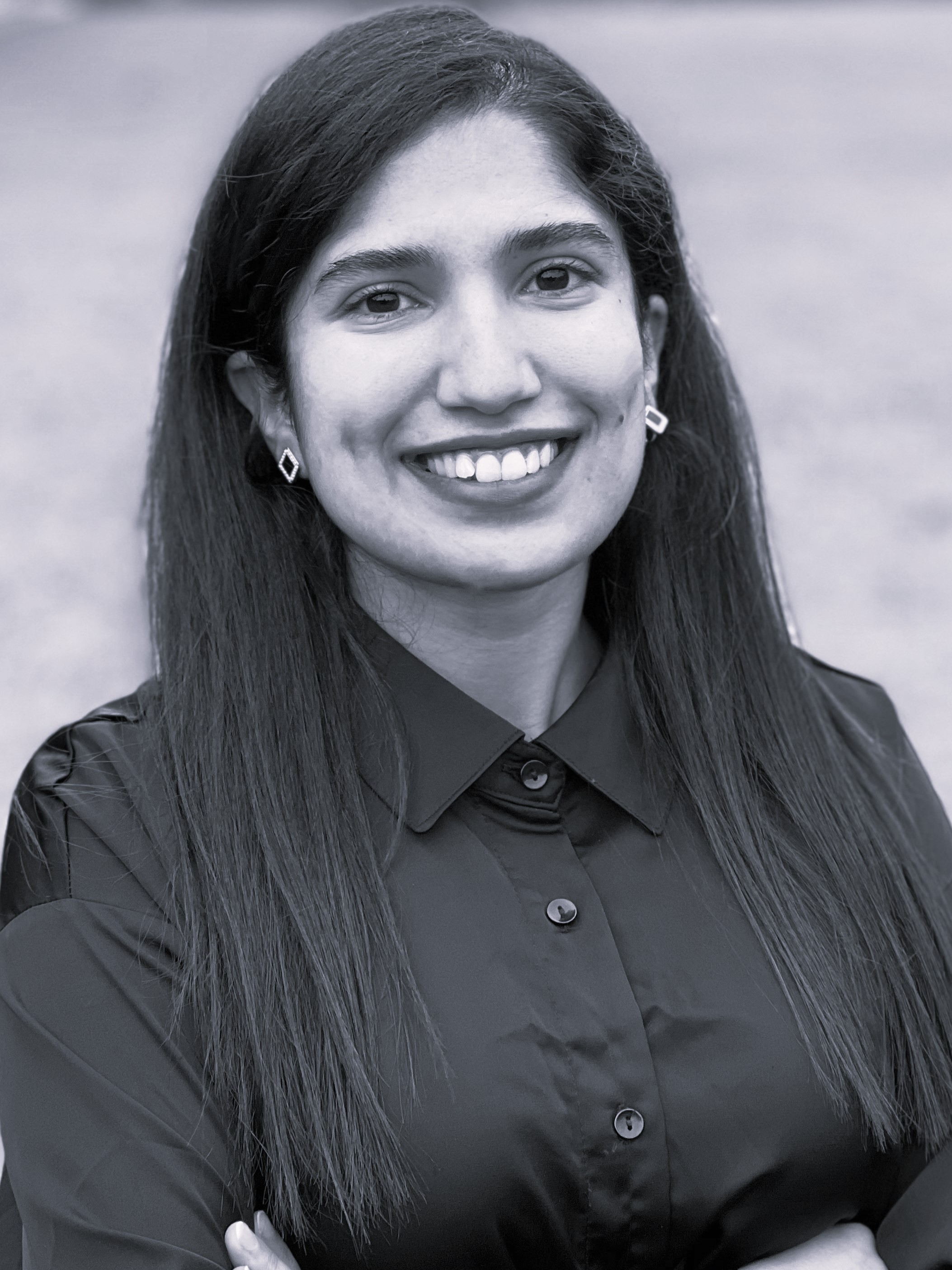}}]{Marziyeh Rezaei}
received her B.S. degree in Electrical Engineering with a focus on Electronic IC Design from Sharif University of Technology, Tehran, Iran, in 2020. She worked as a Research Assistant at the RPI Lab at the Chinese University of Hong Kong during the summer of 2019 and at NVIDIA in the summer of 2024. Currently, she is a Ph.D. student at the University of Washington, Seattle, WA. She was awarded the Cadence Women in Technology Scholarship in 2022. Her research interests include Integrated Circuits and Systems for electro-optical applications.
\end{IEEEbiography}
\vspace{0pt}
\begin{IEEEbiography}[{\includegraphics[width=1in,height=1.25in,clip,keepaspectratio]{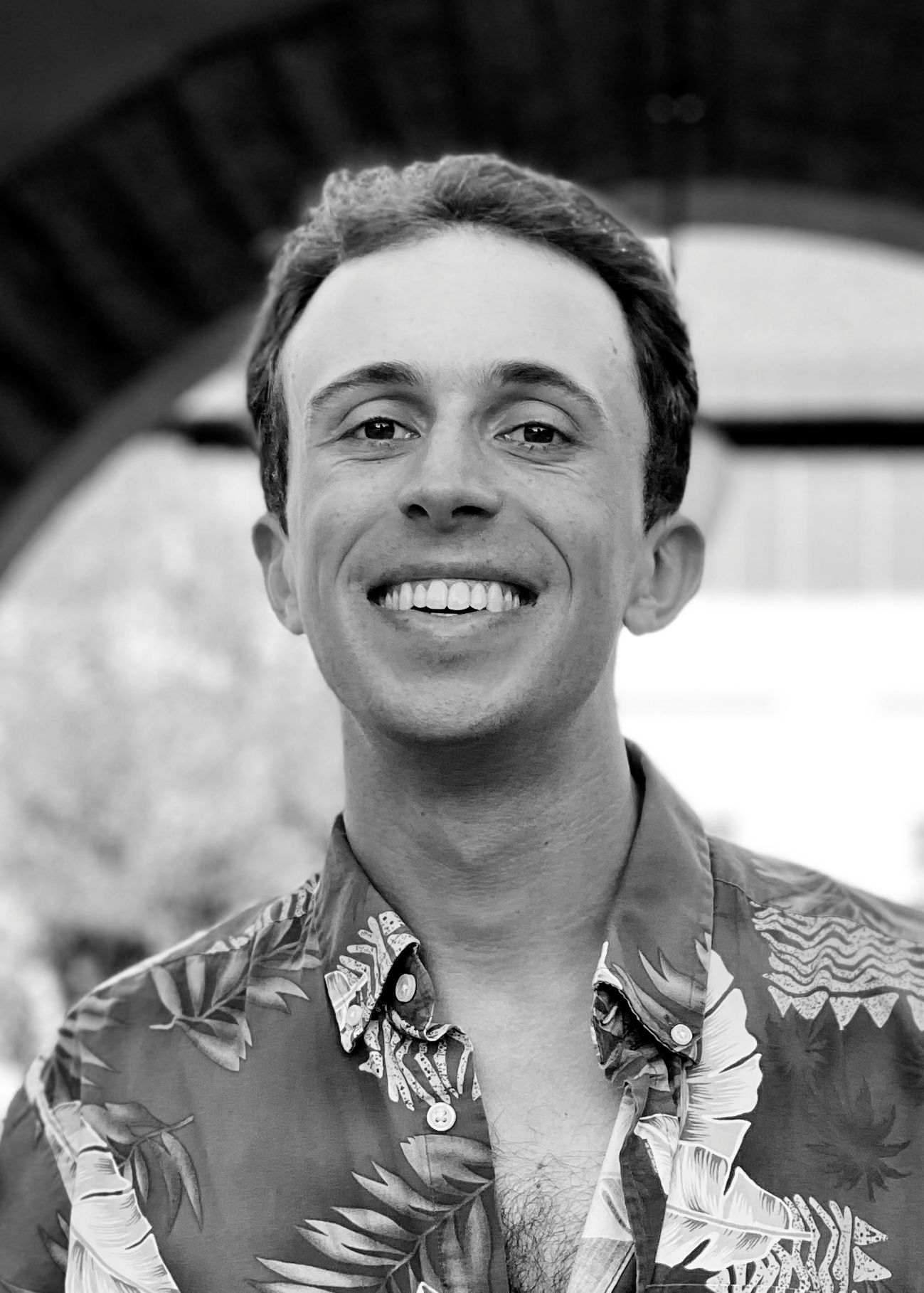}}]{Dan Sturm}
received his bachelors degree (2019) in electrical engineering from Princeton University. He is currently a Ph.D. student at the University of Washington in Seattle, WA. He is a recipient of a three-year fellowship through the NSF Graduate Research Fellowship Program (GRFP) and was previously a recipient of a one-year fellowship from the UW Clean Energy Institute (CEI). His research interests are in photonic integrated circuits, data center interconnect architectures, and electro-optical transceivers. Before coming to UW, Dan worked as an RF systems engineer at Apple and Caltech-based startup GuRu Wireless.
\end{IEEEbiography}
\vspace{-190pt}
\begin{IEEEbiography}[{\includegraphics[width=1in,height=1.25in,clip,keepaspectratio]{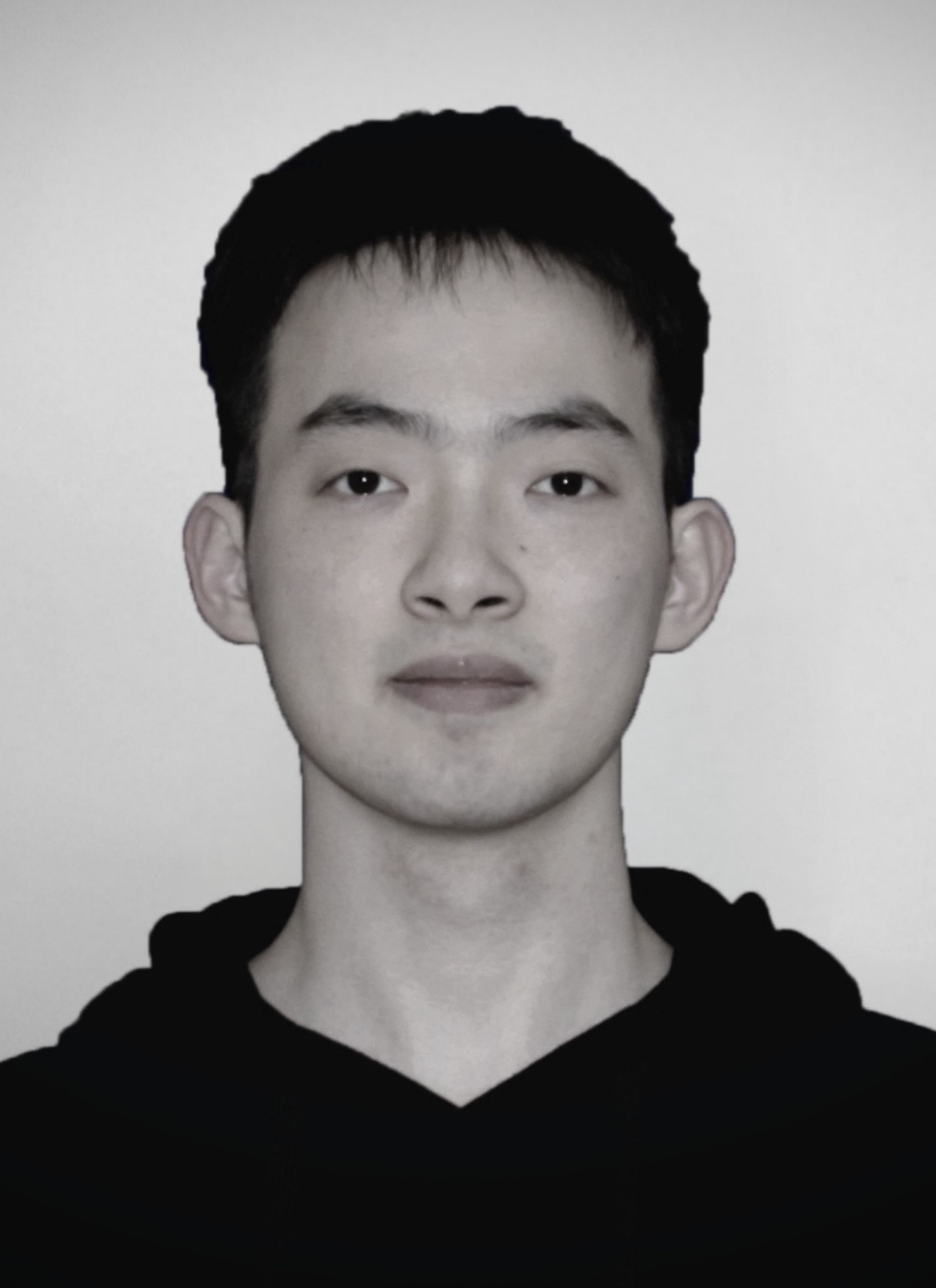}}]{Pengyu Zeng} received his B.Eng. degree in Electronic Information Engineering from Wuhan University, Wuhan, China, in 2023. He is currently pursuing a Ph.D. in Electrical Engineering at the University of Washington, Seattle, WA, USA. In 2022, he served as a research assistant at the University of Notre Dame, Notre Dame, IN, USA. His research interests include analog/mixed-signal and photonic circuits.
\end{IEEEbiography}
\vspace{-190pt}
\begin{IEEEbiography}[{\includegraphics[width=1in,height=1.25in,clip,keepaspectratio]{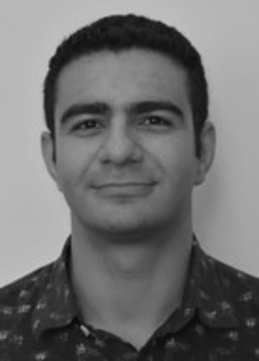}}]{Sajjad Moazeni} received the B.S. degree in electrical engineering from the Sharif University of Technology, Tehran, Iran, in 2013, and the M.S. and Ph.D. degrees in electrical engineering and computer science from the University of California at Berkeley, Berkeley, CA, USA, in 2016 and 2018, respectively.

From 2018 to 2020, he was a Post-Doctoral Research Scientist in Bioelectronic Systems Lab at Columbia University, New York, NY, USA. He is currently an Assistant Professor of Electrical and Computer Engineering Department, at University of Washington, Seattle, WA, USA. He received the 2022 NSF CAREER Award and 2023 Google Faculty Award. His research interests are designing integrated systems using emerging technologies, integrated photonics, neuro and bio photonics, and analog/mixed-signal integrated circuits.
\end{IEEEbiography}
\end{document}